\documentclass[aps,prc,twocolumn,showpacs,floatfix,nobibnotes,superscriptaddress,longbibliography]{revtex4-1}

\usepackage{amsmath}
\usepackage{inputenc}
\usepackage{amssymb}
\usepackage{graphicx}
\usepackage{verbatim}
\usepackage{epsfig}
\usepackage{bm}
\usepackage{color}
\usepackage{float}
\usepackage{dcolumn}
\usepackage{multirow} 
\usepackage[unicode=true,
  linktocpage,
  linkbordercolor={0.5 0.5 1},
  citebordercolor={0.5 1 0.5},
  colorlinks=true,
  linkcolor=blue,
  citecolor=blue,
  urlcolor=blue]{hyperref} 

\usepackage{lipsum} 
\usepackage{placeins} 

\usepackage[normalem]{ulem}
\usepackage[dvipsnames,usenames]{xcolor}

\graphicspath{{.}}

\newcommand{\beq}{\begin{equation}}
\newcommand{\eeq}{\end{equation}}
\newcommand{\beqa}{\begin{eqnarray}}
\newcommand{\eeqa}{\end{eqnarray}}

\newcommand{\NNLOsat}{NNLO$_{\rm sat}$}
\newcommand{\NNLOgod}{$\Delta$NNLO$_{\rm GO}$(450)}
\newcommand{\NLOgod}{$\Delta$NLO$_{\rm GO}$(450)}

\begin{document}

\title{$^{40}$Ca transverse response function from coupled-cluster theory}

\author{J.~E.~Sobczyk}
\affiliation{Institut f\"ur Kernphysik and PRISMA$^+$ Cluster of Excellence, Johannes Gutenberg-Universit\"at, 55128
  Mainz, Germany}

\author{B.~Acharya}
\affiliation{Physics Division, Oak Ridge National Laboratory,
Oak Ridge, TN 37831, USA} 
\affiliation{Institut f\"ur Kernphysik and PRISMA$^+$ Cluster of Excellence, Johannes Gutenberg-Universit\"at, 55128
  Mainz, Germany}
  
\author{S.~Bacca}
\affiliation{Institut f\"ur Kernphysik and PRISMA$^+$ Cluster of Excellence, Johannes Gutenberg-Universit\"at, 55128
  Mainz, Germany}
\affiliation{Helmholtz-Institut Mainz, Johannes Gutenberg-Universit\"at Mainz, D-55099 Mainz, Germany}
  
\author{G.~Hagen}

\affiliation{Physics Division, Oak Ridge National Laboratory,
Oak Ridge, TN 37831, USA} 
\affiliation{Department of Physics and Astronomy, University of Tennessee,
Knoxville, TN 37996, USA}

\begin{abstract}

We present calculations of the $^{40}$Ca transverse response function obtained from coupled-cluster theory used in conjunction with the Lorentz integral transform method. We employ nuclear forces derived at next-to-next-to leading order in chiral effective field theory with and without $\Delta$ degrees of freedom. We first benchmark this approach on the $^4$He nucleus and compare both the transverse sum rule and the response function to earlier calculations based on different methods.
As expected from the power counting of the chiral expansion of electromagnetic currents and from previous studies, our results retaining only one-body term  underestimate  the experimental data for $^4$He by about $20\%$. However, when the method is applied to $^{40}$Ca at the same order of the expansion, response functions  do not lack strength and agree well with the world electron scattering data. We discuss various sources of theoretical uncertainties and comment on the comparison of our results with the available experiments.

\end{abstract}

\maketitle 

\section{Introduction}

A longstanding quest in nuclear physics is to accurately describe a wide range of nuclear phenomena by treating nuclei as systems of protons and neutrons interacting among themselves and with external fields. In a so-called \emph{``ab initio"} approach, the nuclear many-body Schr\"{o}dinger equation is solved, either exactly or using controlled approximations, by employing nuclear forces that are constrained by the symmetries of quantum chromodynamics,  nucleon-nucleon (NN) scattering experiments, and selected nuclear data~\cite{Hergert:2020bxy,ekstrom2023}. Modern \emph{ab initio} computations have increasingly adopted chiral effective field theory ($\chi$EFT) to derive nuclear forces, providing a rigorous framework to organize NN, three-nucleon (3N)  and many-nucleon forces, as well as currents that couple nucleons to external fields, into a low-momentum expansion~\cite{Epelbaum:2008ga}.  The close relation between the nuclear Hamiltonians and electroweak currents makes  $\chi$EFT studies of electroweak observables very interesting~\cite{Gardestig:2006hj, BaccaPastore2014}. Extending  \emph{ab initio} studies to electroweak cross sections is important to test the theory, to estimate theoretical uncertainties, and to possibly improve on the precision.  

Traditionally applied to light and closed-shell medium-mass nuclei, \emph{ab initio} methods have recently found success in describing heavy~\cite{Hu:2021trw}, deformed~\cite{caprio2015,launey2016,novario2020,miyagi2020,frosini2021,hagen2022,ekstrom2023b}, and exotic~\cite{bonaiti2022, Ni68} nuclei.  Coupled-cluster theory~\cite{coester1958,coester1960,kuemmel1978,mihaila2000b, dean2004,wloch2005,hagen2008,hagen2010b, binder2013b,hagen2014} played an important role in these achievements among other computational methods that exhibit a polynomial scaling with the number of nucleons (see Ref.~\cite{Hergert:2020bxy} for a recent review). Extensions of \emph{ab initio} studies to processes that involve nuclear excitations in the continuum is an important frontier~\cite{Johnson:2019sps}.
 Recently, the Lorentz integral transform method~\cite{efros1994} was reformulated within coupled-cluster theory to generate an approach called  LIT-CC which is well suited to study nuclear responses induced by electroweak probes.
Within this approach, the sum over continuous final states is reformulated in terms of an integral transform and later inverted to access the response function.
The LIT-CC method was extensively benchmarked on electric dipole excitations of medium-mass nuclei~\cite{Bacca:2013dma,Bacca:2014rta,Miorelli:2016qbk}, but it can be extended to a variety of electroweak observables.

In addition to improving our understanding of the nuclear dynamics, one of the main motivations to investigate electroweak response functions is the neutrino experiments, especially the long-baseline programs, which require reliable predictions of neutrino-nucleus cross-sections to perform the neutrino energy reconstruction and, from there, infer the oscillations parameters. The quasi-elastic peak is the dominating mechanism in the T2K and the future HyperK~\cite{hyperk} experiments, and is also an important reaction channel for DUNE~\cite{DUNE}.
Presently,  the only \emph{ab initio} predictions of neutrino cross sections in this region are available from the Green's Function Monte Carlo (GFMC) method for light nuclei up to $^{12}$C~\cite{Lovato2014,Lovato:2015qka,Lovato:2017cux,lovato2020}. Under certain approximations, other methods, such as the short-time approximation~\cite{STA} or the spectral function~\cite{Rocco:2018vbf, Barbieri:2019ual,Sobczyk:2020qtw}, have been proposed to connect the cross sections with \emph{ab initio}  calculations. Recently there has also been progress in computing the $^{16}$O spectral function calculated within coupled-cluster theory~\cite{Asia2023}.
 
In this endeavor, electron-scattering experiments play an important role in guiding theoretical studies~\cite{CLAS:2021neh}. Due to an analogous structure of electron- and neutrino-nucleus cross-sections, the precise electron data provide a stringent test for our modeling of the nuclear dynamics.
The Rosenbluth technique allows a separation of the electron-scattering cross section into the longitudinal and the transverse response functions, which are driven by the electric charge and the electromagnetic current operators, respectively. Using the LIT-CC approach, we successfully described the longitudinal response function at the momentum transfers of a few hundred MeV in the region medium-mass nuclei~\cite{Sobczyk:2020qtw,Sobczyk:2021dwm}. The final state interactions were consistently taken into account, leading to a very good agreement with available experimental data on $^4$He and $^{40}$Ca.

In this work, we further extend the LIT-CC method to describe the transverse response function, which dominates the cross section at backward scattering angles. We provide the first \emph{ab initio} results for mass number $A=40$ in the impulse approximation. This is an important step towards building a complete and systematic theory for lepton-nucleus scattering.
 
The article is structured as follows. In Sec.~\ref{sec:formalism}, we introduce the definition of the transverse response function and describe the framework in which we perform calculations. We present our results in Sec.~\ref{sec:results}, starting from the transverse sum rule for $^4$He and $^{40}$Ca and benchmark with existing calculations. Next, we present the response functions for these two nuclei up to $400$ MeV/c, commenting on the tensions existing in available experimental data on $^{40}$Ca. Finally, we conclude in Sec.~\ref{sec:conclusions}.

\section{Formalism}
\label{sec:formalism}
The inclusive electron-nucleus scattering cross-section in the Born approximation can be expressed in terms of longitudinal and transverse response functions. The focus of this work is the transverse response function $R_T$, which depends on the energy and momentum transfer $(\omega,q)$, and is defined as 
\beq
\label{eq:rt_def}
R_T(\omega,q)\!=\!\int \!\!\!\!\!\!\!\sum _{f} 
|\!\left\langle \Psi_{f}| \mathbf{J}^T (q)| 
\Psi _{0}\right\rangle\!|
^{2}\,\delta\!\!\left(\!E_{f}+\frac{q^2}{2M}-E_{0}-\omega \! \right)\,.
\eeq
Here, $\mathbf{J}^T$ corresponds to the transverse part of the electromagnetic current operator with respect to the direction of the vector momentum transfer ${\mathbf q}$. In Eq.~({\ref{eq:rt_def}}), the nuclear ground state wavefunction is denoted by $\Psi _{0}$ (with corresponding energy $E_0$) and we sum over all possible final states $\Psi _{f}$ (with energy $E_f$) which obey the energy conservation expressed by the delta function, where $M$ is the mass of the nucleus and $q=|{\mathbf q}|$.

In the impulse approximation, the current operator can be decomposed into the isoscalar and isovector parts $ \mathbf{J} =  \mathbf{J}_{s} +  \mathbf{J}_{v}$, as
\begin{align}
\label{eq:isovectorjvec}
& \mathbf{J}_{s} =\sum_{j=1}^A \left( G_E^S(Q^2)\frac{\mathbf{{p}}_j}{m} 
   - i \, G_M^S(Q^2)\frac{\mathbf{q}\times\bm{\sigma}_j}{2m} \right) e^{i\mathbf{q}\cdot\mathbf{r}_j}\frac{1}{2} \,, \nonumber \\
& \mathbf{J}_{v} =\sum_{j=1}^A \left( G_E^V(Q^2)\frac{\mathbf{{p}}_j}{m} 
   - i \, G_M^V(Q^2)\frac{\mathbf{q}\times\bm{\sigma}_j}{2m} \right) e^{i\mathbf{q}\cdot\mathbf{r}_j}\frac{\tau^{z}{_j}}{2} \,.
\end{align}
Here, one can clearly identify the convection and spin terms, which depend on the nucleon momenta ${\mathbf p}_j$ and spins $\bm{\sigma}_j$, respectively. 
The nucleon mass is indicated with $m$ and its third isospin component with $\tau^{z}_j$, while $G_{E,M}^S \equiv  G_{E,M}^p+G_{E,M}^n$ , $ G_{E,M}^V \equiv G_{E,M}^p-G_{E,M}^n $ are
 the electric and magnetic form factors related to the proton and neutron form factors.
They are functions of $Q^2=q^2-\omega^2$. However, in our calculation the dependence on the energy transfer $\omega$ is not known \emph{a priori}, so we assume $\omega\approx\sqrt{q^2+m^2}-m$ which corresponds to the kinematical region of the quasi-elastic peak. This approximation has been previously applied in GFMC calculations~\cite{lovato2013, lovato2020}. 
While several parameterizations exist for the form factor, in our calculations we use that from Ref.~\cite{Kelly:2004hm}. 

The transverse response function is known to receive a large contribution from  two-nucleon currents which are required by the continuity equation~\cite{BaccaPastore2014}. This fact has been confirmed by Green function Monte Carlo calculations for light nuclei up to $^{12}$C using  phenomenological potentials and currents~\cite{lovato2020}. It is also expected from the power counting when performing the chiral expansion of electromagnetic currents, where the two-body currents appear at the same order as the leading one-body currents in the counting advocated by Refs.~\cite{Kolling:2009iq,Krebs:2020pii} and are suppressed by only one chiral order in the power counting adopted by Refs.~\cite{Pastore:2008ui,Acharya:2021lrv}. In this work we only include the one-body current contributions as a first step. The study of the role of two-body currents is left for future work. 

\subsection{Lorentz integral transform}

\label{sec:lit}
The sum over continuum spectrum of final states in Eq.~\eqref{eq:rt_def} poses a serious computational challenge. 
To overcome it, we follow the Lorentz integral transform method~\cite{efros2007}. First, we calculate the convolution of the nuclear response function with the Lorentzian kernel 
\begin{equation}
K(\omega, \sigma) = \frac{\sigma_I}{\pi}\frac{1}{(\omega-\sigma_R)^{2}+\sigma_I^{2}}\, 
\end{equation}
as 
\begin{equation}
{\cal L}_T(\sigma,q)= \int d\omega R_T(\omega,q) K(\omega, \sigma)\, ,
\label{eq:lit}
\end{equation}
where  $\sigma = \sigma_R + i \sigma_I$ is a complex parameter.
Using Eqs.~\eqref{eq:rt_def} and \eqref{eq:lit}, ${\cal L}_T(\sigma,q)$ can be formally expressed as the solution of a ``Schr\"{o}dinger-like'' equation with a source term 
\begin{equation}
\label{lit}
(H-E_{0}-\sigma)|\widetilde{\Psi}_{\sigma}(q) \rangle=\mathbf{J}^T(q)|{\Psi_{0}}\rangle\,.
\end{equation}
The latter can be solved with a good many-body bound-state technique, so that the integral transform of the response function becomes
\begin{equation}
{\cal L}_T(\sigma,q)= \langle\widetilde{\Psi}_{\sigma}(q)
|\widetilde{\Psi}_{\sigma}(q) \rangle 
\label{liteq}
\end{equation}
and can be numerically evaluated.
The response $R_T(\omega, q)$ is then retrieved from ${\cal L}_T(\sigma,q)$ through an inversion procedure.
In general, the inversion is an ill-posed problem which requires some additional assumptions about the shape of the response function~\cite{efros2007}.
We invert the transform assuming that the final response function is smooth. This assumption is fully justified in the region of the quasi-elastic peak.

\subsection{Coupled-cluster method}

In the coupled-cluster method many-body correlations are systematically included to the wave function through an exponential ansatz $e^T$ acting on a reference state $\Phi_0$. The correlation operator $T$ is a linear expansion in $n$-particle $n$-hole excitation operators with respect to the reference state, 
\begin{equation}
  T=T_1+T_2+\ldots +T_A 
\end{equation}
The Schr\"{o}dinger equation, rewritten with the coupled-cluster ansatz, becomes then
\begin{equation}
    H_N e^T |\Phi_0\rangle = E_0 e^T|\Phi_0\rangle\ \ \ \rightarrow \ \ \\  \overline{H}_N  |\Phi_0\rangle = E_0 |\Phi_0\rangle\, ,
\end{equation}
where $\overline{H}_N \equiv e^{-T} H_N e^T$ 
is the similarity transformed normal ordered Hamiltonian, and $\Phi_0$ is the Hartree-Fock solution to a Hamiltonian containing nucleon-nucleon and three-nucleon forces. The excitation operator $T$ is usually truncated at some low excitation rank. 
In this work, we use $T \approx T_1 + T_2$, which is known as coupled-cluster with single and double excitations (CCSD)~\cite{shavittbartlett2009}. This approximation is known to account for $\sim 90\%$ of the correlation energy in the ground state for systems with a well defined Fermi surface~\cite{bartlett2007}.

\subsection{The LIT-CC method}

 The Lorentz integral transform method has been recently formulated within coupled-cluster theory~\cite{bacca2013}. In this language, Eq.~(\ref{liteq}) becomes
\begin{align}
\nonumber
    &(\overline{H}_N-E_{0}-\sigma)|\widetilde{\Psi}_{\sigma}^R(q) \rangle= \overline{\Theta}_J |{\Phi_{0}}\rangle\,, \\
    &\langle \widetilde{\Psi}_{\sigma}^L(q)|  (\overline{H}_N-E_{0}-\sigma) = \langle \Phi_{0}|(1+\Lambda )\,\overline{\Theta}_J ,
\label{eom}
\end{align}
where $\overline{\Theta}_J \equiv e^{-T} \Theta_J e^T$  is  a similarity transformed operator.
Given that coupled-cluster theory is non-Hermitian, we solve both the left and right equations
separately for $\widetilde{\Psi}_{\sigma}^L(q)$ and $\widetilde{\Psi}_{\sigma}^R(q)$ to get 
\begin{equation}
{\cal L}_T(\sigma,q)= \langle
\widetilde{\Psi}_{\sigma}^L(q)
|\widetilde{\Psi}^R_{\sigma}(q) 
\rangle\,.
\label{lit_cc}
\end{equation}
The operator $\Lambda=\Lambda_1+\Lambda_2+...$ is a de-excitation operator which has an analogous structure as $T$ operator and accounts for the fact that the left and right eigenstates are not the same. 
The equations in (\ref{eom}) are both equations-of-motion with a source term. They can be solved,  after having solved for the ground-state, using similar  techniques to the standard ones to get the spectrum of excited states within coupled-cluster theory~\cite{stanton1993,bacca2013}.  

The operator $\Theta_J$  is a multipole, i.e., a spherical rank-$J$ tensor, of the transverse electromagnetic current, namely $\mathbf{J}^T(q) \equiv \sum_{J=0}^{J_{max}} \Theta_J$. 
Expressions for the multipoles of the current operators can be found in Ref.~\cite{Sonia_ee'}.
After calculating the LITs of Eq.~(\ref{lit_cc}) for each rank-$J$ multipole, we sum them up, and finally perform an inversion of the total integral  transform.

\section{Results}
\label{sec:results}

We now present results focusing on the nuclei $^4$He, which we will use as a benchmark, and $^{40}$Ca.  
All the results presented in this section are obtained within a model space of 15 major oscillator shells which proved to be sufficient for systems and observables we compute~\cite{Sobczyk:2021dwm}. An additional cut is imposed on the three-body potential $e_{3{\rm max}} =   2n_1 + l_1  + 2n_2 + l_2 + 2n_3 + l_3  \le 16$. We check the convergence of the model space by varying the underlying harmonic oscillator frequency. 
We use two different chiral Hamiltonians at the next-to-next-to-leading order: \NNLOsat\ \cite{Ekstrom:2015rta} and \NNLOgod~\cite{jiang2020}. In the latter one, the $\Delta$ degree of freedom is considered explicitly in the construction of nuclear potential. Both interactions have the same regulator cut-off of $450$ MeV/c. They predict correct binding energies and radii for light and medium-mass nuclei, in particular the ones we consider.

\subsection{Transverse sum rule} 

The transverse sum rule is defined in a standard way as
\begin{align}
\label{eq:tsr}
    \text{TSR}(q) = \frac{2m^2}{Z\mu_p^2+N\mu_n^2}\frac{1}{q^2} \Big( & \langle \Psi_0| \mathbf{J}^{T\dagger} \mathbf{J}^T |\Psi_0\rangle\nonumber \\ 
    &-| \langle \Psi_0| \mathbf{J}^T |\Psi_0\rangle |^2  \Big)\,.
\end{align}
with the nucleon magnetic moments $\mu_{n,p}$.
The elastic contribution (second line) is zero for the closed-shell nuclei which we consider. The normalization is chosen in such a way that $\text{TSR}(q\to\infty  ) \approx 1$. We use this limit as a numerical check of the multipole decomposition of the current. 
Let us first benchmark our transverse sum rule result for $^4$He. We employed the \NNLOsat\ potential with the underlying harmonic oscillator frequency $\hbar\Omega=16$ MeV, in accordance with our previous benchmarks of the longitudinal response and Coulomb sum rule~\cite{Sobczyk:2020qtw}.  

\begin{figure}[hbt]
	\includegraphics[width=0.4\textwidth]{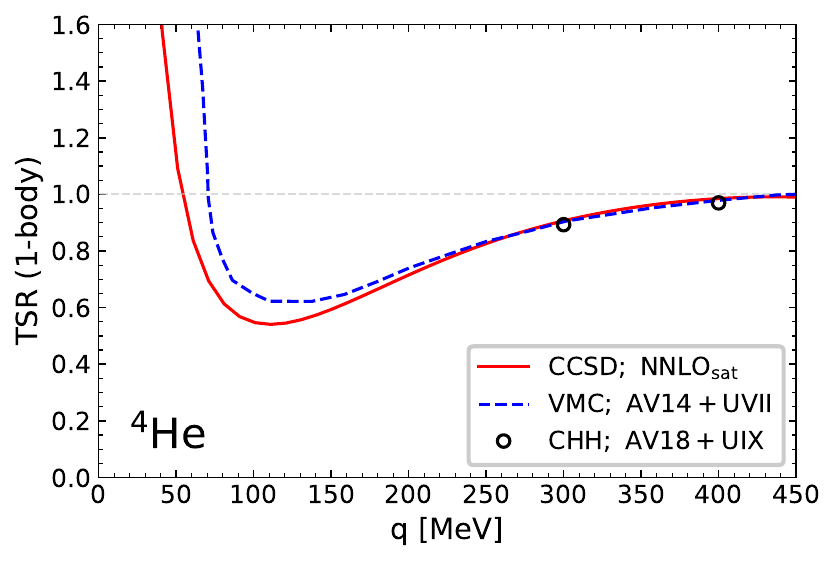}
	\caption{$^{4}$He results for transverse sum rule for N2LO$_{\mathrm{sat}}$  and $\hbar \Omega=16$ MeV compared with  results of Refs.~\cite{Schiavilla:1988ff,Carlson:2001mp}.}
	\label{fig:TSR_4He}
\end{figure}

As shown in Fig.~\ref{fig:TSR_4He}, we obtain a very good agreement above $q=200$ MeV/c when comparing with other simulations which use either AV14+UVII~\cite{Schiavilla:1988ff} or AV18+UIX potentials~\cite{Carlson:2001mp}, for the variational Monte Carlo (VMC) and the correlated hyperspherical harmonics (CHH) methods, respectively. The low-$q$ behavior diverges (because of the $1/q^2$ factor in the definition of the sum rule, see Eq.~\eqref{eq:tsr}) and differs quite substantially from the calculation of Ref.~\cite{Schiavilla:1988ff} that employed a phenomenological potential. In the limit of $q\to0$, the spin part of the current becomes negligible, while the convection-current contribution becomes proportional to the kinetic energy (see Eq.~\eqref{eq:isovectorjvec}). The latter strongly depends on the employed Hamiltonian, which explains the difference between the calculations. We performed an internal check of our method, comparing our result with an average kinetic energy $\langle T\rangle = \int d^3p n(p) p^2/2m$ where $n(p)$ is the momentum distribution. We obtained agreement within a few percent. 

\begin{figure}[hbt]
	\includegraphics[width=0.4\textwidth]{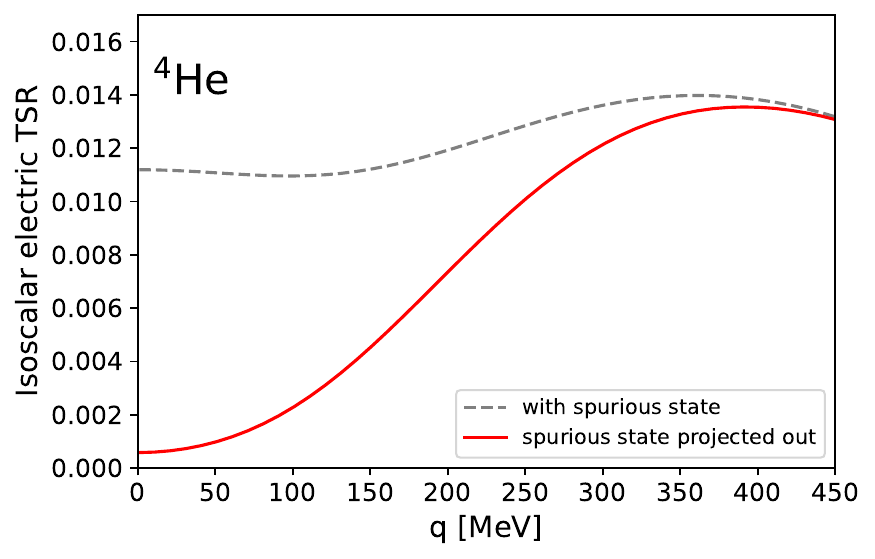}
	\caption{The effect of spurious states removal in $^{4}$He for the isoscalar electric part of the current.}
	\label{fig:TSR_isoscEl}
\end{figure}

In our previous works we investigated the role played by spurious center-of-mass states excited by the charge operator and devised a procedure to remove them. We do not repeat the procedure here, but rather refer the reader to Ref.~\cite{Sobczyk:2020qtw} for details.
Instead, we present here the result of an analogous analysis of center of mass effects performed on the transverse sum rule. We checked the various contributions of the transverse current operator and found out that,  as expected, only the isoscalar electric part suffers from substantial spurious contributions. The difference between a calculation with spurious states and one where those are projected out is shown in Fig.~\ref{fig:TSR_isoscEl}. Moreover, this part of the current is an order of magnitude smaller compared to the isovector one. We conclude that the  spurious-states contamination plays here a negligible role. 

\begin{figure}[hbt]
	\includegraphics[width=0.4\textwidth]{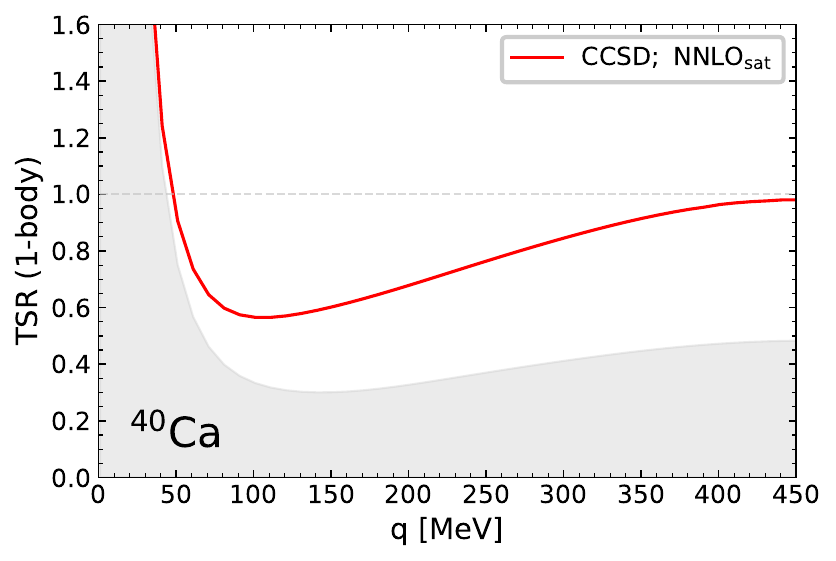}
	\caption{$^{40}$Ca results for the transverse sum rule. We employ the \NNLOsat, the model space of 15 major oscillator shells  and $\hbar \Omega=22$ MeV. The shadowed region shows the contribution of the electric multipoles only (without the magnetic multipoles).}
	\label{fig:TSR_40Ca}
\end{figure}

\begin{figure}[hbt]
	\includegraphics[width=0.5\textwidth]{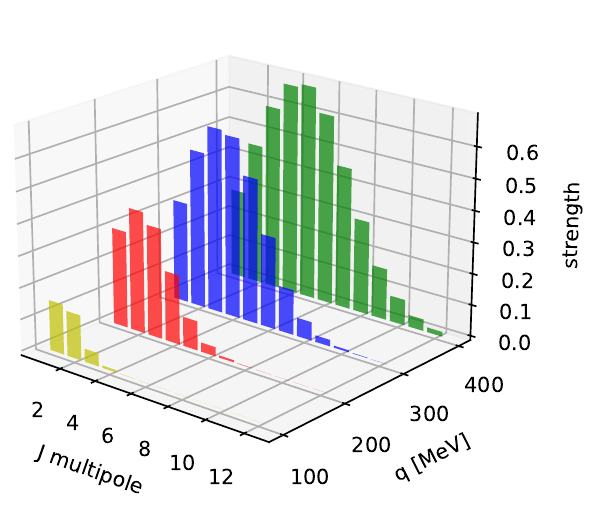}
	\caption{The strength of various multipoles of the decomposed response as a function of momentum transfer $q$ for $^{40}$Ca. }
	\label{fig:multipoles_40Ca}
\end{figure}

Finally, in Fig.~\ref{fig:TSR_40Ca} we show the TSR result for $^{40}$Ca obtained with the \NNLOsat\ potential using the harmonic oscillator frequency $\hbar\Omega=22$ MeV. Its behaviour is qualitatively similar to $^4$He. We also present separately the electric contribution (shadowed area) which dominates over the magnetic one at lower $q$. 
The number of multipoles needed to converge and their relative strength depends on the size of the nucleus and the momentum transfer. In Fig.~\ref{fig:multipoles_40Ca}, we show this dependence for four values of the momentum transfer, $q=100-400$ MeV/c in case of $^{40}$Ca. This information is useful to estimate the numerical cost for calculating the response functions. The number of multipoles needed to get a convergent result grows from four for $q=100$ MeV/c up to twelve for $q=400$ MeV/c (for $^4$He these are three and six, respectively). 

\subsection{Transverse Response functions}

We now turn our attention to the transverse response function, which we calculate from the inversions of the LITs, using an ansatz for the response function.
We expand it in terms of the following basis functions
\begin{align}
    &R_T(\omega, q) = \sum_{i=1}^N c_i f_i(\omega; n_0,\beta_i)\,, \\ \nonumber
    &f_i(\omega; n_0,\beta_i) = (\omega-\omega_{th})^{n_0} e^{-\frac{ (\omega-\omega_{th})}{\beta_i}}\,.
\end{align}
This form of basis functions is motivated by the threshold behavior of the response function for the deuteron which is known analytically to be $\omega^{\ell+1/2}$ ($\ell$ is the orbital angular momentum quantum number). The threshold energy $\omega_{th}$ is nucleus-dependent and in our case corresponds to the proton separation energy. 
For various values of $\sigma_I$, $n_0$ and $N$, we vary the continuous parameter $\beta_i$ over a wide range. For each combination, we search for the optimal $c_i$ coefficients minimizing the least square fit.

For all the results in this section we performed a robust analysis, varying $n_0=\{0.5,1.5\}$, and the number of basis functions $N=\{6,7,8\}$. We repeated the simulation for two values of  $\sigma_I$, namely 5 and 10 MeV.  We estimate the inversion uncertainty as the spread between all of the found optimal solutions.

\begin{figure}[thb]
    \includegraphics[width=0.4\textwidth]{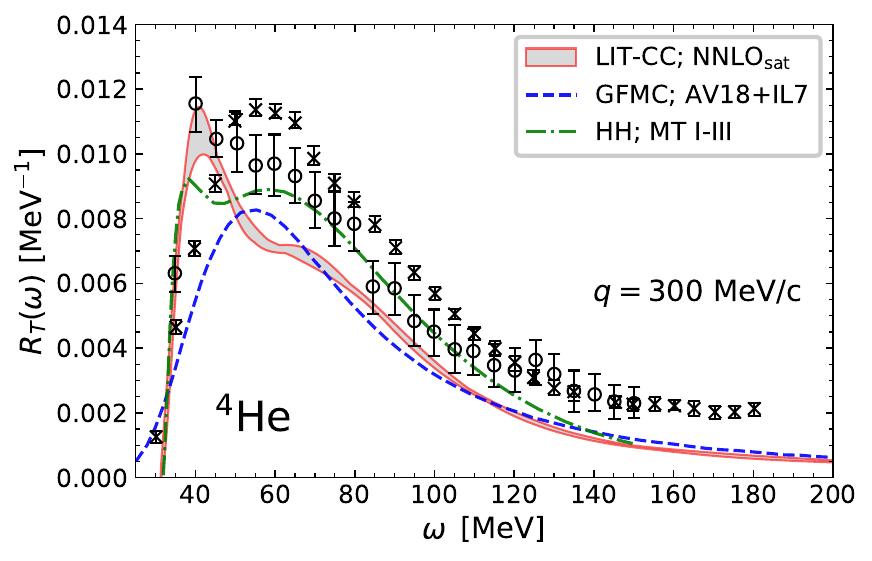}
    \includegraphics[width=0.4\textwidth]{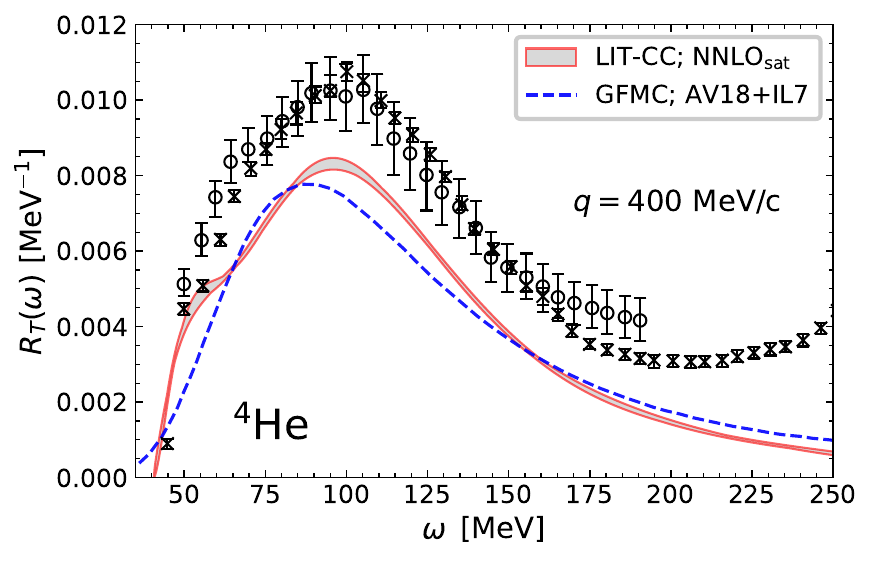}
	\caption{ $^{4}$He transverse response function  for $q=300$ and  $400$ MeV/c using the \NNLOsat\  interaction compared with GFMC predictions using AV18+IL7 potential~\cite{Rocco:2018tes} and hyperspherical harmonics with the MTI-III interaction~\cite{Bacca:2006ji}. Experimental data are taken from Refs.~\cite{Dytman:1988fi} (circles) and \cite{Zghiche:1993xg} (crosses). }
	\label{fig:4He}
\end{figure}

In Fig.~\ref{fig:4He}, we present our benchmark results for $^4$He at the kinematics of $q=300$ and $400$ MeV/c, comparing them with experimental data and other theoretical calculations obtained with one-body currents only. The uncertainty band of our CCSD results comes from the inversion procedure, not from a variation of the harmonic oscillation frequency, which is kept fixed at $\hbar \omega $=16 MeV, given that varying it yields negligible differences.  The band is a few percent broad and  becomes of the order of $10\%$ only in the lower part of the spectrum for $q=300$ MeV/c. 
For both considered momentum transfers we observe around $20\%$ missing strength with respect to the experimental data. This fact was already predicted using other nuclear Hamiltonians and many-body methods and was attributed to the lack of two-body currents~\cite{Lovato:2015qka}. 

\begin{figure}[thb]
    \includegraphics[width=0.4\textwidth]{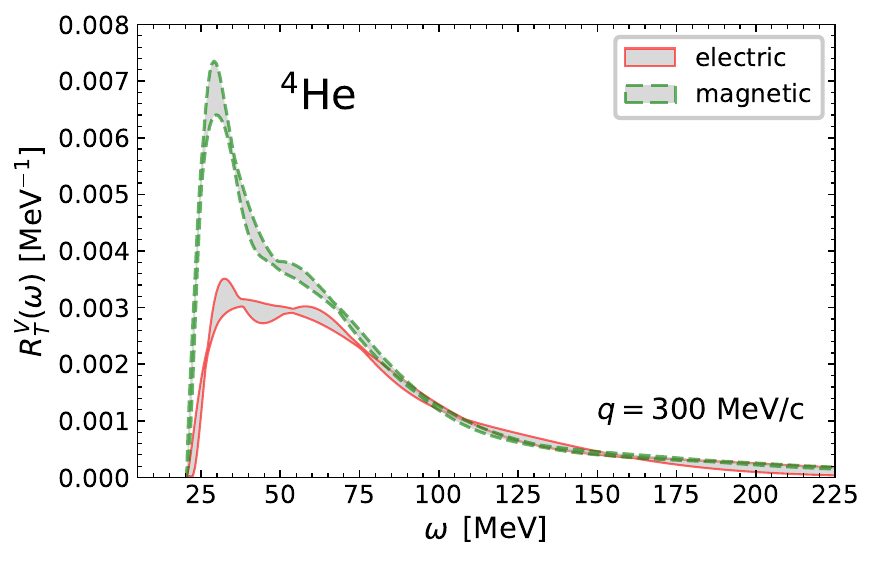}
    \includegraphics[width=0.4\textwidth]{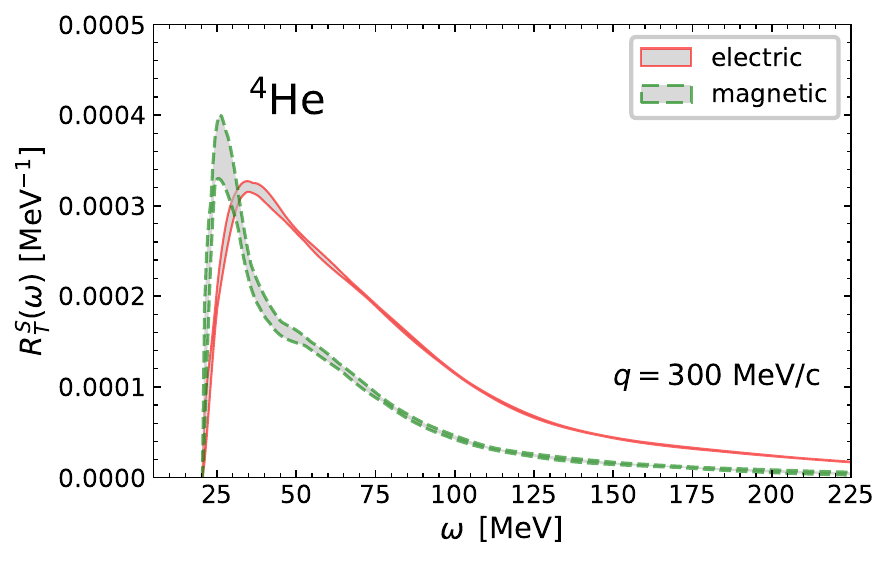}    
	\caption{Response functions on $^4$He at $q=300$ MeV/c. The isovector (upper panel) and isoscalar (lower panel) contributions are shown separately for the electric and magnetic part of the transverse current.}
	\label{fig:4He_contributions}
\end{figure}

We observe, especially for $q=300$ MeV/c (also visible for $q=400$ MeV/c), an irregular shape of the quasi-elastic peak. The visible two-peak structure is qualitatively similar to the one obtained from hyperspherical harmonics calculations with the MTI-III potential. For the GFMC result we observe a much smoother quasi-elastic peak with a different low-energy behaviour. We interpret these differences as a result of using $(i)$ different integral transforms: the Lorentz (in our case) or Laplace (for the GFMC), $(ii)$ different nuclear forces,  as well as $(iii)$ different inversion procedures. We notice that a similar discrepancy has been already reported for the longitudinal response function in Ref.~\cite{Rocco:2018tes} where the GFMC results were compared with the hypersherical harmonics using the same interaction but different integral transforms.
To get more insight about the shape of the response function, in Fig.~\ref{fig:4He_contributions} we split the contributions to the transverse response function at $q=300$ MeV/c into the electric/magnetic and isoscalar/isovector parts. This reveals that the the low-energy shoulder originates from the magnetic part of current and that the isovector contribution is an order of magnitude larger than the isoscalar.

We proceed to the calculation of the transverse response function for $^{40}$Ca. 
In this case, we use two different chiral Hamiltonians, employed previously for the longitudinal response function~\cite{Sobczyk:2021dwm}, namely  the \NNLOsat\ ~\cite{Ekstrom:2015rta} and \NNLOgod~\cite{jiang2020}  chiral potentials at  next-to-next-to-leading order in a $\chi$EFT with and without $\Delta$ degrees of freedom, respectively. 
For the inversion, we use the theoretical values of the proton separation energy, namely $\omega_{th}=6.9$ MeV for \NNLOsat\ and $\omega_{th}=8$ MeV for \NNLOgod, imposing  the response function  to be zero below this threshold. In our studies of the longitudinal response function we found  that the dependence on the $\hbar\Omega$ is negligible when varying it between 16 and 22 MeV. Here, we performed an analogous analysis using $\hbar\Omega=16$ and $22$ MeV and we arrived at the same conclusion for the transverse response function.

Let us briefly comment on the available experimental data for $^{40}$Ca. Two experiments performed the Rosenbluth separation in the past~\cite{Williamson:1997zz,Meziani:1984is} and their results are in tension. In particular, the authors of Ref.~\cite{Meziani:1984is} found a $25-40\%$ lower Coulomb sum rule. Later, a re-analysis of the world data was performed by Jourdan~\cite{Jourdan:1996np}, who presented the results for three values of momentum transfer $q=300,\, 380,$ and $570$ MeV/c. However, there is not a single value of momentum transfer, for which the data of all of these three analyses is available. Therefore, we decided to show our results at several kinematics against various data-sets.

First we look at $q=300$ and $380$ MeV/c, for which the world data is available. For $300$ MeV/c, the re-analysis by Jourdan overlaps to great extent with the original data points from Ref.~\cite{Williamson:1997zz}, apart from a few points above 120 MeV. A very good agreement between our calculation and the data for both kinematics is observed, although somehow unexpectedly, since for light nuclei the one-body current gives only $70-80\%$ of the strength with respect to the data.

\begin{figure}[thb]
    \includegraphics[width=0.4\textwidth]{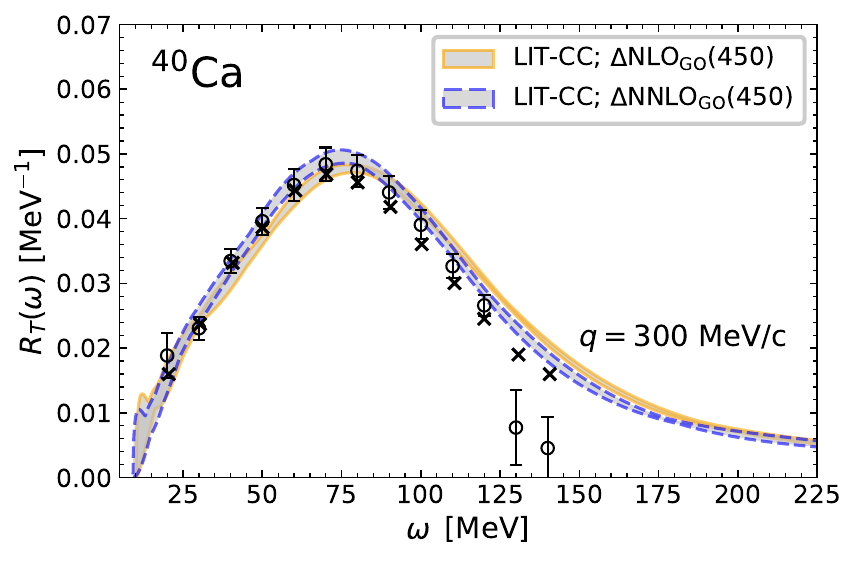}
	\caption{Dependence of the transverse response function on the chiral order of the $\Delta$-full interaction. Experimental data aretaken from~\cite{Williamson:1997zz} (circles) and ~\cite{Jourdan:1996np} (crosses).}
	\label{40Ca-chiralOrder}
\end{figure}

Within our framework we are able to assess several sources of uncertainty, both coming from the chiral expansion of the employed Hamiltonian, as well as from the truncations in the many-body method. 
To estimate systematic uncertainties from $\chi$EFT, we check the response at $q=300$ MeV/c, for which we used potentials at two orders: \NLOgod\ and \NNLOgod\ of the chiral Hamiltonian (see Fig.~\ref{40Ca-chiralOrder}).
The difference between them is at the order of a few percent. This is analogous to our findings for the longitudinal response function~\cite{Acharya:2022drl}. We leave the further studies of the convergence of the many-body method for the future, since it will require much more expensive calculations including triple correlations. However, we can say that first pilot runs including triples excitations do not show any sizable difference in the quasi-elastic peak. It remains to be seen what the effect of two-body currents will be. Work to include them is presently ongoing~\cite{Acharya}. 

\begin{figure}[thb]
    \includegraphics[width=0.4\textwidth]{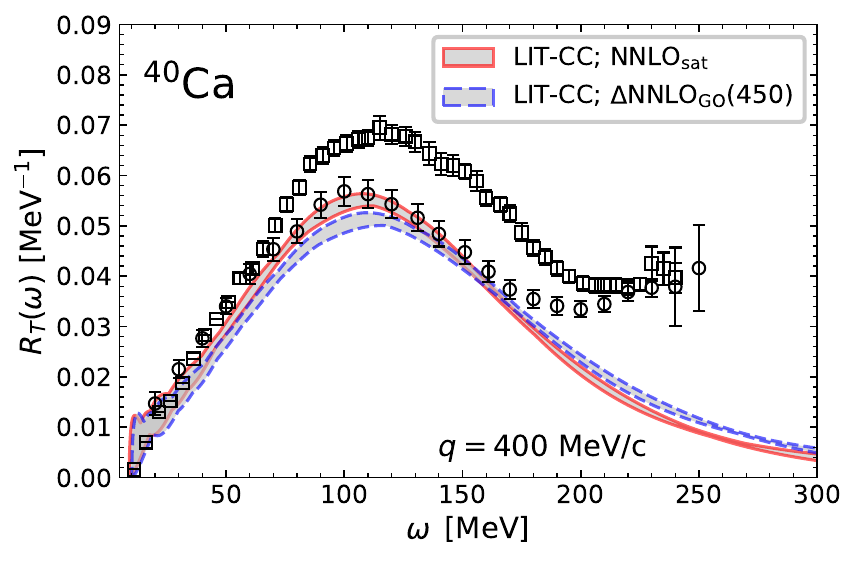}
    \includegraphics[width=0.4\textwidth]{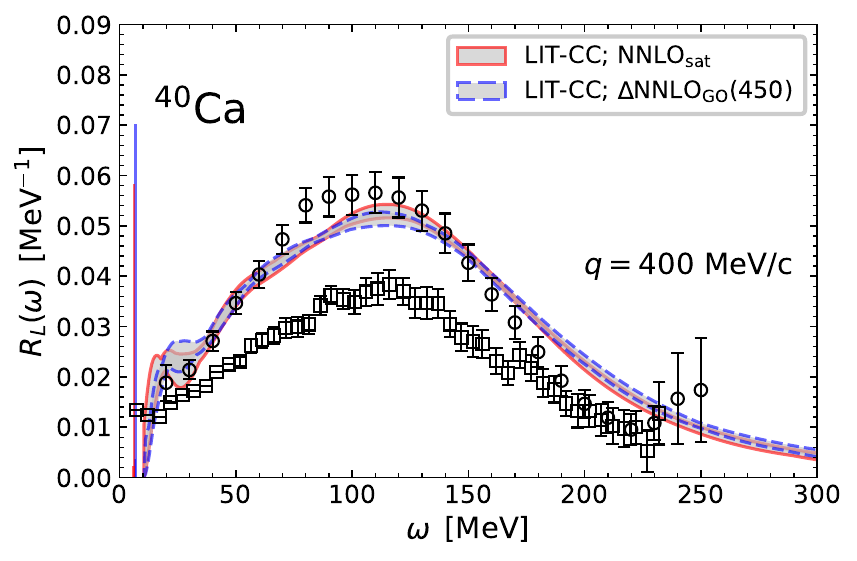}
	\caption{Transverse and longitudinal response functions for $^{40}$Ca calculated for $q=400$,\ MeV/c for \NNLOsat\ and \NNLOgod\ potentials. Experimental data for $q=400$ MeV/c taken from~\cite{Williamson:1997zz} (circles) and for $q=410$ MeV/c from~\cite{Meziani:1984is} (squares). }
	\label{40Ca-2datasets}
\end{figure}

Finally, to illustrate the tension between the experiments, in Fig.~\ref{40Ca-2datasets} we show data for $q=400$ MeV/c~\cite{Williamson:1997zz} and $q=410$ MeV/c~\cite{Meziani:1984is} and compare them with our calculations done at $q=400$ MeV/c, both for the longitudinal and transverse response functions. Experimental data-sets differ even by $30\%$ in the quasielastic peak, although a $10$ MeV difference between momentum transfers is not expected to give such a big effect (rather of the order of a few percent). Our results agree considerably better with Ref.~\cite{Williamson:1997zz}. However, the larger prediction of $R_T$ by Ref.~\cite{Meziani:1984is}, would be more consistent with the assumption of large two-body currents contribution. 

\begin{figure}[thb]
    \includegraphics[width=0.4\textwidth]{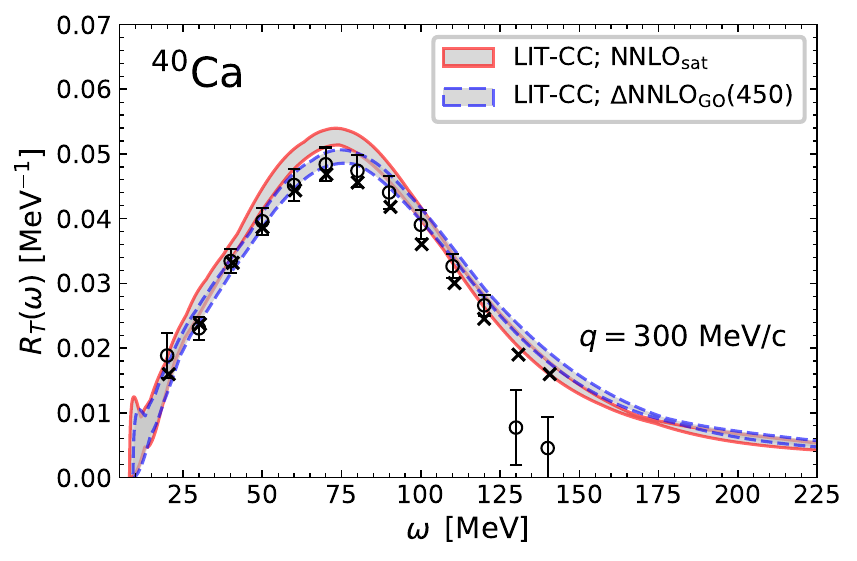}
    \includegraphics[width=0.4\textwidth]{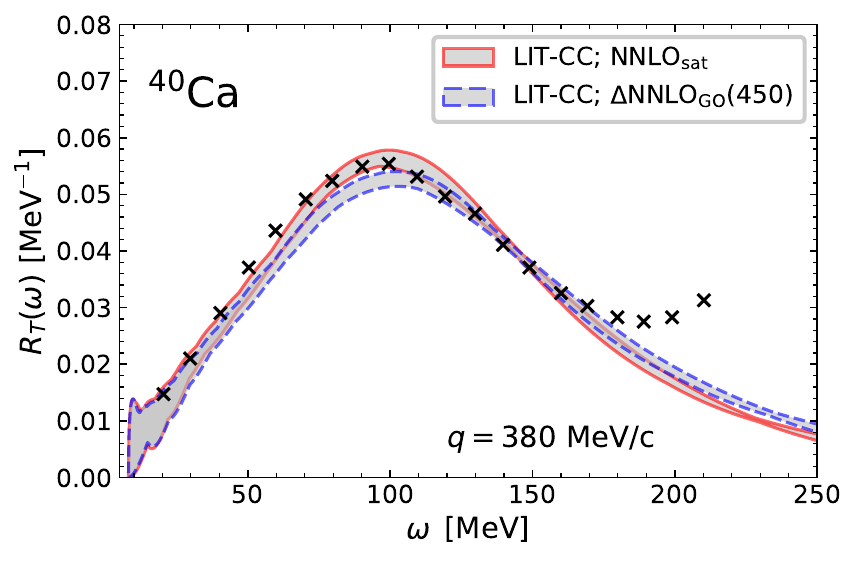}
	\caption{$^{40}$Ca transverse response function  for $q=300$ and $380$ MeV/c with the \NNLOsat\ and \NNLOgod\ potentials. Experimental data are taken from~\cite{Williamson:1997zz} (circles) and ~\cite{Jourdan:1996np} (crosses). }
	\label{40Ca}
\end{figure}

\section{Conclusions}
\label{sec:conclusions}
We presented the first \emph{ab initio} calculation of the transverse response in the region of medium-mass nuclei computed within the LIT-CC method. 
This work is another step forward in the development of our theoretical program towards calculating electroweak responses for the future neutrino oscillation experiments. 

We benchmarked our method on $^4$He where we found a reasonable agreement with previous calculations and experimental data. Our results underestimate the data, which is expected since we included only one-body current. Nevertheless, the transverse sum rule as a function of $q$ has a very similar behavior as other calculations performed with other methods and phenomenological potentials. 

Next, we performed an analysis of the transverse response function on $^{40}$Ca. Our results stay in a very good agreement with the world data by Jourdan. 
Our finding motivates further investigations: both the inclusion of the two-body currents, as well as the study of truncations in the many-body expansion are to be performed in the future.
We also want to point out that the tension between the `two experimental Rosenbluth data on $^{40}$Ca have not yet been fully understood and clarified. In view of the long-baseline neutrino programs and of the role  electron scattering  plays in constraining  theoretical predictions, 
it is important to shed light on this matter. A new experimental program in this direction is being initiated in Mainz~\cite{Doria}.

\begin{acknowledgments}
This project has received funding from the European Union’s Horizon 2020 research and innovation programme under the Marie Skłodowska-Curie grant agreement No.~101026014.
This work was supported by the Deutsche
Forschungsgemeinschaft (DFG) 
through the Cluster of Excellence ``Precision Physics, Fundamental
Interactions, and Structure of Matter" (PRISMA$^+$ EXC 2118/1) funded by the
DFG within the German Excellence Strategy (Project ID 39083149).
It is also supported by the U.S. Department of Energy, Office of Science, Office of Nuclear Physics under Award Number DE-SC0018223 (SciDAC-4 NUCLEI)  and the SciDAC-5 NUCLEI collaboration, and by
the Office of High Energy Physics, U.S. Department of Energy, under
Contract No. DE-AC02-07CH11359 through the Neutrino Theory Network Fellowship awarded to BA. Computer time was provided by the supercomputer MogonII at Johannes Gutenberg-Universit\"{a}t Mainz, and by the Innovative and Novel Computational Impact on Theory and Experiment (INCITE) programme. This research used resources of the Oak Ridge Leadership
Computing Facility located at Oak Ridge National Laboratory, which is
supported by the Office of Science of the Department of Energy under
contract No. DE-AC05-00OR22725 
\end{acknowledgments}

\FloatBarrier 
\bibliography{master,refs} 

\end{document}